\documentclass[aps, pre, twocolumn, showpacs, groupedaddress]{revtex4-2}

\usepackage[french]{babel}
\usepackage[utf8]{inputenc}
\usepackage[T1]{fontenc}
\usepackage{amsthm}
\usepackage{amsmath,amssymb,amsfonts}
\usepackage{mathtools}
\usepackage{physics}
\usepackage{xcolor}
\usepackage{graphicx}
\usepackage{subcaption}
\usepackage[normalem]{ulem}
\usepackage{array}

\begin{document}
\title{Intruder dynamics in granular media under localized surface loading}

\author{E. M. Franklin$^{1,2}$}
\author{B. Darbois Texier$^1$}
\author{A. Seguin$^1$}
\author{D. D. Carvalho$^{1,2}$}
\author{Y. Bertho$^{1}$}

\affiliation{$^1$ Universit\'e Paris-Saclay, CNRS, Laboratoire FAST, 91405 Orsay, France}
\affiliation{$^2$ Faculdade de Engenharia Mec\^anica, Universidade Estadual de Campinas (UNICAMP), Rua Mendeleyev, 200 Campinas, SP, Brazil}

\begin{abstract}
We experimentally investigate the dynamics of a spherical intruder driven horizontally at a constant force in a granular medium subjected to a localized surface overload. While intruder motion beneath a free surface exhibits constant acceleration in the quasistatic regime, the presence of a surface load induces a pronounced transient deceleration when the intruder passes below the loaded region. The magnitude of this deceleration increases with the applied overload and saturates at large overloads, while it decreases with intruder depth. Introducing a characteristic timescale and an overload-based Froude number, we show that the deceleration dynamics collapse onto master curves. We further develop a model incorporating stress transmission from the surface, which partially captures the intruder deceleration. In this approach, this deceleration is shown to depend on two parameters: the overload and the area on which this overload is applied. These results provide a framework to quantify how localized surface stresses influence subsurface intruder dynamics, with implications for locomotion, root growth, and underground transport in granular media.
\end{abstract}

\maketitle

\section{Introduction}

From everyday experience, we know that slowly displacing a solid object (intruder) within a granular medium is more difficult than in a fluid, as it involves stronger force fluctuations, intermittency, and vibrations. For example, it is much easier to stir a teaspoon in a cup of tea than in a container filled with granulated sugar. In the latter case, the discrete nature of grains creates and breaks contact chains that intermittently resist and collapse under the imposed force, resulting in a much more complex motion of the intruder. In spite of its complexity, this kind of motion is frequent in nature, such as in the displacement of animals \cite{hosoi2015beneath} and the growth of plant roots \cite{kolb2017physical} within sand, and in human activities as well, such as soil drilling \cite{feng2022granular} or the construction of foundations in civil engineering. Therefore, improving our understanding of such motions is essential both for natural systems and for human activities.

Numerous studies have investigated the quasistatic motion of intruders within granular media, providing a clear characterization of the drag forces they experience \cite{Albert, Albert2, Stone, Geng, Costantino, Hilton, Kozlowski, Carlevaro, Kolb1, Seguin1, Carvalho4}. The drag force on an intruder is observed to scale with its cross-sectional area multiplied by the granular pressure at its location, which in turn increases linearly with depth \cite{Brzinski, Rodriguez}. In this regime, the drag force is found to be independent of the intruder  velocity \cite{Seguin1, Seguin3, Andreotti_6, Carvalho, Carvalho4}, in contrast to what is typically observed in fluids. Furthermore, experimental and computational studies have investigated the rearrangement of grains as an intruder moves through a granular medium \cite{Kolb1, Seguin1}. These studies show that grains reorganize in a localized region surrounding the intruder, becoming compressed at the front and dilated in the trailing region, which can result in local jamming. Nonetheless, most of these studies focus on the motion of an intruder beneath a free granular surface, where surface loads are absent.

The influence of a load applied to the surface of a granular medium has been primarily investigated in static configurations. For a localized point load, the resulting stress distribution has been shown to follow a Gaussian profile, which broadens and weakens with increasing depth in the packing \cite{reydellet2001green,bouchaud2002stress,geng2003green}. When the load covers the entire surface of a granular column filling a tube, most of the overload is screened after a characteristic distance on the order of the tube diameter \cite{cambau2013local}, in agreement with Janssen’s effect \cite{bertho2003}. However, the impact of such loads on dynamic processes has received  far less attention. This raises several natural questions: how deeply do overload-induced stresses propagate into the bed, and does their influence saturate as the added load increases? More broadly, does an animal moving within sand or a root growing in soil feel the overload caused by a given weight placed on the bed surface? If the stress signature indeed extends into the bulk and can be quantitatively related to the applied overload, this could pave the way to practical applications. Yet, these questions remain largely unexplored.

In this paper, we investigate how an additional vertical load applied to the surface of a granular bed affects the quasistatic motion of a spherical intruder moving horizontally beneath the surface. To this end, we carried out experiments in which a constant driving force was applied to the intruder, while its displacement was recorded over time at varying depths and surface loads. Across all tested conditions, we systematically observed a deceleration as the intruder passed beneath the overloaded surface, with a magnitude that depends on the surface load. This deceleration was found to saturate as the overload increased. Additionally, we observed that the proximity of the walls can significantly influence the drag force experienced by the intruder. Based on the intruder’s equations of motion, we introduce a characteristic timescale, which, together with a Froude number accounting for the overload, provides an appropriate normalization for the problem. Finally, we present an analytical model that captures the deceleration and can be used to estimate the displacement of intruders in loaded granular media.

\section{Experimental setup}

\begin{figure*}[t]
    \centering
    \includegraphics[width=0.75\textwidth]{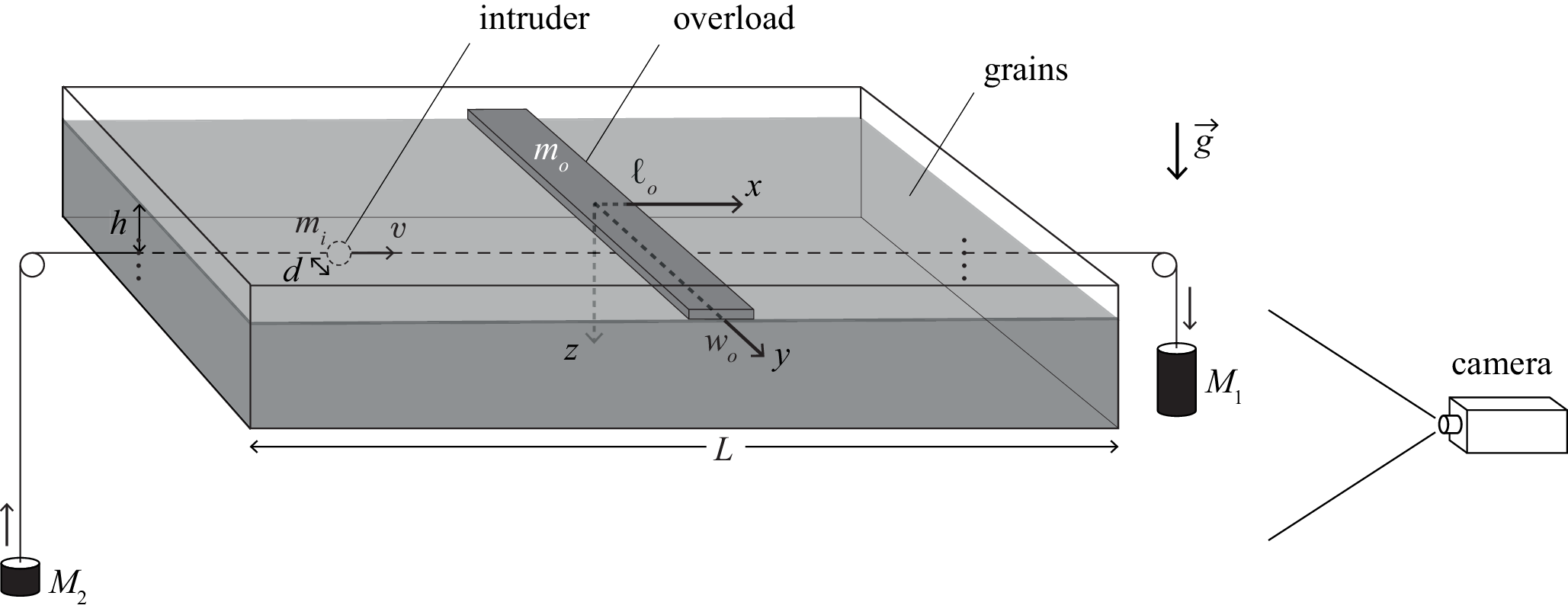}
    \caption{Sketch of the experimental setup used to study the motion of a spherical intruder of diameter $d$ immersed in a granular medium and driven at velocity $v$ by the mass difference $M_1-M_2$. The setup allows investigation of the effect on the intruder dynamics of an overload of mass $m_o$ applied on a surface $w_o\ell_o$ placed over the granular bed.}
    \label{Fig01}
\end{figure*}

The experimental setup consists of a rectangular container measuring $L=364$~mm long ($x$ direction) and 264~mm wide ($y$ direction), filled with monodisperse glass beads of diameter $d_g=1\pm 0.1$~mm up to a height of 96~mm (Fig.~\ref{Fig01}). 
The glass beads of density 2500~kg~m$^{-3}$ yield a granular medium with a bulk density of approximately $\rho\approx 1600$~kg~m$^{-3}$. To ensure reproducible experimental conditions, the granular medium is stirred between successive experiments and subsequently leveled by gently shaking the container horizontally, yielding a flat free surface. The origin of the coordinate system is defined at the center of the container, at the level of the granular surface. A spherical intruder of mass $m_i=2.1$~g and diameter $d=12$~mm is initially immersed in the granular bed, near one end of the container. Four small holes (diameter smaller than 0.5~mm) are drilled into each of the two vertical walls perpendicular to the longitudinal direction, at depths $h=36, 46, 56, \mathrm{and\ } 66$~mm below the free surface. These openings allow a nylon thread, attached to the intruder, to pass through the container at a controlled depth. The container rests on a horizontal table, and the nylon line exits horizontally before being redirected vertically downward by pulleys located at the edge of the table. Two suspended masses, $M_1$ and $M_2$, are attached to the ends of the thread, thereby exerting a constant horizontal driving force $\Delta Mg = (M_1-M_2)g$ on the intruder, where $g$ is the gravitational acceleration. The vertical components of the tension compensate both the intruder's weight and the lift force acting on it \cite{Guillard}, preventing any significant vertical displacement. The mass $M_1$ is varied between 140.6 and 300.3~g, while $M_2$ is fixed at 50.7~g, resulting in driving forces ranging from about 0.9 to 2.5~N. Because the intruder and the masses are connected by the same nylon thread, we assume that their displacements along the horizontal $x$ and vertical $z$ directions, respectively, remain approximately identical throughout the motion. As direct visual access to the intruder is not possible, the motion of mass $M_1$ is tracked using a high-speed camera (resolution $2560\times 1600$~pixels) operating at a frame rate of 500~Hz, corresponding to a spatial resolution of approximately 4 pixels per millimeter. The acquired images are subsequently postprocessed to extract the trajectory of $M_1$, from which the intruder dynamics are inferred. The intruder velocity is obtained by differentiating the position signal in time, using a central-difference scheme combined with moving-average smoothing. The position data are first smoothed over 20 frames (0.04~s), followed by time differentiation, and the resulting velocity signal is further smoothed over 30 frames (0.06~s). For all investigated depths, the ratio $d/h$ ranges from 0.18 to 0.33. A slight and transient deformation of the free surface is observed only at the shallowest depth ($h=36$~mm, $d/h \approx 0.33$), with a maximum amplitude of about 2~mm that relaxes after the passage of the intruder. No significant surface deformation is detected at larger depths, indicating that the present study primarily focuses on a regime where free-surface effects remain limited.

Finally, to investigate the effect of localized surface loading on the intruder dynamics, a rectangular plate is placed on the top of the granular bed, centered on the container to ensure a symmetrical overload. The plate extends $w_o=55$~mm along the $x$ direction and $\ell_o=264$~mm along the $y$ direction, thereby spanning the entire width of the container, as shown in Fig.~\ref{Fig01}. The applied overload mass $m_o$ is varied between 29.1 and 2123.1~g.

Table~\ref{Tab01} summarizes the experimental conditions, corresponding to an applied surface stress $\sigma_o=m_og/w_o\ell_o$ ranging from 20 to 1500~Pa. Note that each experiment is repeated six times to compute mean values and statistically meaningful standard deviations.\\

\begin{table}[h]
	\centering
	\begin{tabular}{| @{\hspace{10pt}} c @{\hspace{10pt}} | @{\hspace{10pt}} c @{\hspace{10pt}} | @{\hspace{10pt}} c @{\hspace{10pt}} |}
		\hline
		$h$ & $M_1$ & $m_o$\\
		(mm) & (g) & (g)\\
		\hline
		36	&	140.6	&	0 -- 600.0\\
		46	&	171.6	&	0 -- 250.0\\
		56	&	200.7	&	0 -- 900.0\\
		66	&	300.3	&	0 -- 2123.1\\
		\hline
	\end{tabular}
	\caption{Summary of the experimental conditions: immersion depth $h$ of the intruder, mass $M_1$ setting the horizontal driving force, and overload mass $m_o$ applied at the surface of the granular bed. The mass $M_2$ is fixed at 50.7~g.}
	\label{Tab01}
\end{table}

\section{Results and discussion}

\subsection{Intruder dynamics in an unloaded granular medium}
\label{subsection_without_overload}

\begin{figure*}[t]
    \centering
    \includegraphics[width=\textwidth]{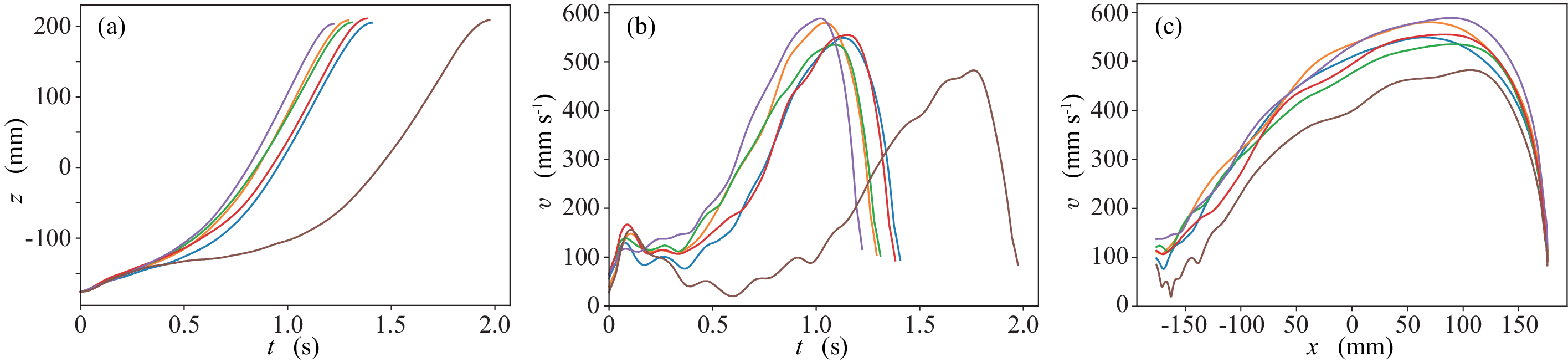}
    \caption{(a)~Position $z$ and (b)~velocity $v$ of the mass $M_1$ as a function of time $t$. (c)~Velocity $v$ of the intruder as a function of its position $x$ in the container. The different curves correspond to different trials performed in the same experimental conditions, with $\Delta Mg\simeq 1.2$~N and at an immersion depth $h=46$~mm.}
    \label{Fig02}
\end{figure*}

We first investigate the horizontal displacement of the intruder through the grains, without any overload on the surface, i.e., $m_o =0$. Figures~\ref{Fig02}(a) and \ref{Fig02}(b) show, respectively, the vertical position $z$ and the corresponding velocity $v$ of the mass $M_1$ as a function of time $t$, for six equivalent experimental runs performed under identical conditions, at an immersion depth $h=46$~mm and with a driving force $\Delta Mg\simeq 1.2$~N.
We observe in Fig.~\ref{Fig02}(a) that the mass undergoes accelerated motion over time. While the onset time of the motion shows significant variability, the overall dynamics remain highly reproducible across different tests. As shown in Fig.~\ref{Fig02}(b), the velocity initially displays a bump at the onset of motion, followed by small oscillations. It then increases approximately linearly with time, before undergoing a pronounced deceleration at the end of the trajectory, when the intruder reaches the container wall. The linear increase of velocity with time indicates that the system undergoes constant acceleration.

Regarding motion onset, the variability observed in Fig.~\ref{Fig02}(a) is attributed to the intruder’s proximity to the container wall, which hinders grain recirculation around it and thereby delays the initiation of movement. At the end of the trajectory, as the intruder approaches the opposite wall of the container, the rapid deceleration arises from grain rearrangement around the intruder, which becomes increasingly constrained near the boundary and thus enhances resistance to motion. The wall itself ultimately stops the intruder.

Analyzing more precisely the total distance traveled by the falling mass in Fig.~\ref{Fig02}(a), we note that it is slightly greater than the container length, by up to 20~mm (maximum observed value). This discrepancy is attributed to the slight elasticity of the nylon thread connecting the intruder to the driving mass. In the initial moments, the thread stretches before transmitting sufficient force—a behavior supported by the early-time velocity curves, which show a small bump associated with thread relaxation [see Fig.~\ref{Fig02}(b)]. In the following, we assume that the wire extension occurs only during the very first moments of the mass dynamics, and we subtract this extension in order to deduce the intruder position in the container from the mass position. Figure~\ref{Fig02}(c) shows the intruder velocity $v$ as a function of its position $x$, corrected for wire extension. The total distance traveled by the intruder then matches the container length, i.e., 352~mm.

These experiments were repeated for different driving forces $\Delta Mg$ and intruder depths $h$. In all trials, a regime was consistently observed in which the intruder undergoes constant acceleration. Naturally, both the intruder velocity and acceleration depend on its depth of immersion and on the driving force applied. For a given immersion depth, a minimum driving force is required to initiate motion. This threshold increases with depth, reflecting the greater resistance offered by the granular medium as the intruder is  deeper in the bed. Above the threshold, the intruder’s acceleration increases with the driving force $\Delta Mg$.

To rationalize these observations, we consider the equation of motion of the spherical intruder. The intruder experiences a drag force from the granular medium, whose form depends on the flow regime, characterized by the Froude number $\mathrm{Fr} = v / \sqrt{gh}$, defined in terms of the intruder velocity $v$ and depth $h$.
Throughout the experiments, the Froude number remains strictly below 1 in nearly all cases, except for about 10 out of more than 300. In these conditions, the granular flow surrounding the intruder remains quasistatic, and the inertial contribution of the drag force it experiences can be neglected. The drag originates from the vertical stress generated by the weight of the grains, which scales as $\rho gh$, acting on the intruder frontal area $\pi d^2/4$, and is expressed as $K_z d^2\rho gh$, where $K_z$ is a numerical coefficient \cite{Guo}. Since the density of the intruder is comparable to that of the granular medium, an added mass force must be taken into account, as is commonly done in fluid mechanics. The added mass $m_a$ corresponding to this force reads $m_{a}=C_{a}\pi\rho d^3/6$, where $C_{a}\simeq 1.24$ is a coefficient that depends on the grain size $d_g$ and the diameter of the intruder $d$ \cite{Seguin4}. Consequently, the total mass to be considered in the equation of motion is $m=m_i+M_1+M_2+m_a$. Defining $a_\mathrm{qs}=d^2x/dt^2$ as the acceleration of the intruder in the regime where the quasistatic drag force dominates, this acceleration reads
\begin{equation}
      a_\mathrm{qs} = \frac{\Delta M - K_z d^2\rho h}{m} g,
    \label{eq_a_char}
\end{equation}
from which, by considering $h$ as the typical length, we obtain the characteristic time $\tau=\left(h/{a_\mathrm{qs}}\right)^{1/2}$.
By normalizing the position $x$ by $h$ and time $t$ by $\tau$, and considering the initial condition $x(t= 0)= 0$, we obtain the expression for the intruder's velocity $v$ in its normalized form
\begin{equation}
    \frac{v}{v_c} = \frac{t}{\tau} ,
    \label{eq_motion_normaliz}
\end{equation}
where $v_c=h/\tau$ is the characteristic velocity of the problem. Equation~(\ref{eq_motion_normaliz}) predicts that the intruder velocity increases linearly with time, in qualitative agreement with the experiments. To assess the quantitative agreement between the model and the data, it is necessary to estimate $\tau$, which depends on the coefficient $K_z$. In the literature, $K_z$ has been reported to vary widely, with values ranging from 1 to 8 \cite{Katsuragi2, Seguin5, Guo, Carvalho3}.
\begin{figure}[t]
    \centering
    \includegraphics[width=\columnwidth]{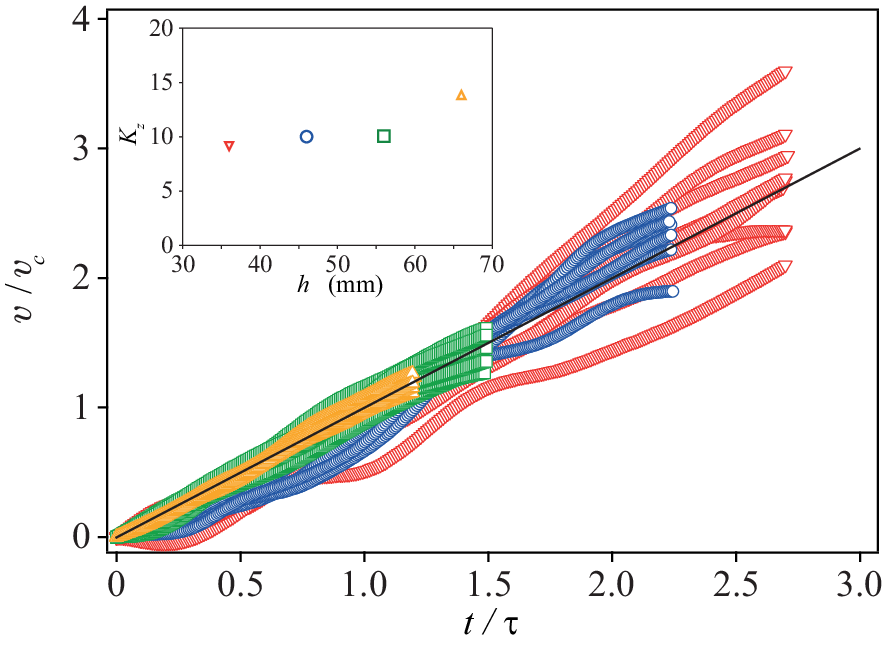}
    \caption{Linear phase of $v$ normalized by the characteristic velocity $v_c$, as a function of the normalized time $t/\tau$, for the depths (\textcolor{red}{$\triangledown$})~$h=36$~mm, (\textcolor{blue}{$\circ$})~$h=46$~mm, (\textcolor{green}{$\square$})~$h=56$~mm, and (\textcolor{orange}{$\vartriangle$})~$h=66$~mm. (---)~Straight line with slope 1, and passes through the origin. Inset: coefficient $K_z$ as a function of the immersion depth $h$. Standard deviations are too small to be shown.}
    \label{Fig03}
\end{figure}
While Guo proposed that $K_z$ primarily depends on the packing fraction \cite{Guo}, a wide range of variability may still be expected due to other influencing factors such as surface roughness of the intruder and the grains, relative humidity, polydispersity, and other system properties \cite{Katsuragi}. Therefore, we fit the experimental data using Eq.~(\ref{eq_motion_normaliz}), treating $K_z$ as a free parameter. Figure~\ref{Fig03} shows the normalized intruder speed $v/v_c$ as a function of the normalized time $t/\tau$ and a reasonable collapse of the data on the unitary slope is observed. The values of the coefficient $K_z$ obtained from this fitting procedure are displayed in the inset of Fig.~\ref{Fig03}. We find that $K_z$ remains approximately constant across the first three immersion depths, with $K_z\simeq 10$, but increases to $K_z\simeq 14$ for the deepest configuration ($h=66$~mm). This increase is attributed to the proximity of the intruder to the bottom wall, as the available distance below the intruder is reduced to approximately 2.5 intruder diameters. In the quasistatic regime, grain rearrangements responsible for the drag remain strongly localized around the intruder \cite{Seguin3}. When these localized rearrangements are affected by the bottom boundary, the grains can no longer reorganize freely beneath the intruder, leading to an increased resistance to motion and therefore to a larger effective drag coefficient. This explains why the deepest configuration systematically departs from the behavior observed at the smaller immersion depths throughout the manuscript. This reference behavior in the absence of surface loading will serve as a baseline to quantify the effects of an applied overload in the following section.

\subsection{Intruder dynamics in a loaded granular medium}
\label{subsection_with_overload}

\subsubsection{Experimental results}
\begin{figure*}[t]
    \centering
    \includegraphics[width=\textwidth]{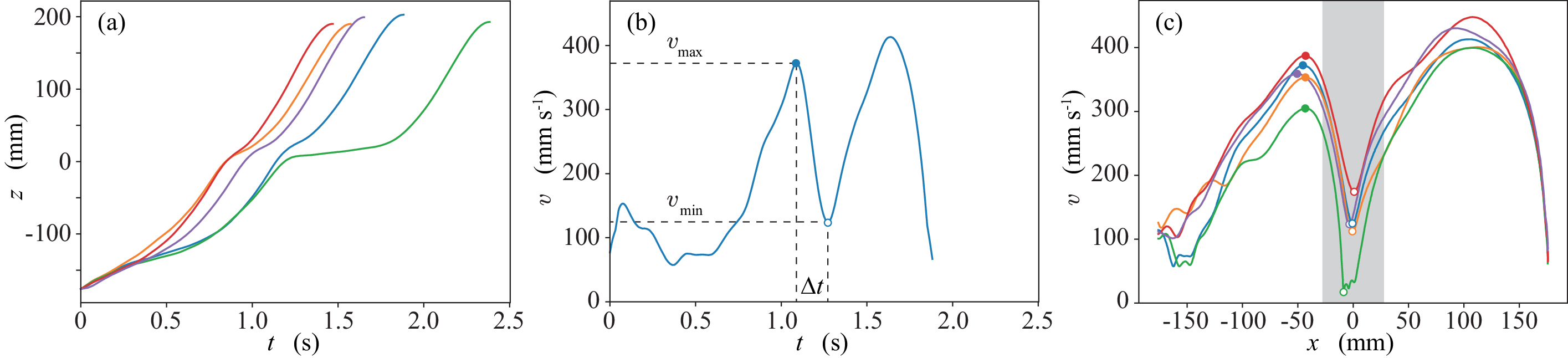}
    \caption{(a)~Position $z$ and (b)~velocity $v$ of the mass $M_1$ as a function of time $t$. (c)~Velocity $v$ of the intruder as a function of its position $x$ in the container. Filled circles indicate the point where deceleration due to the surface overload becomes noticeable, while open circles mark the end of the overload's influence zone. The shaded area indicates the position of the surface overload. The different curves correspond to different trials performed in the same experimental conditions, with an overload of mass $m_o=223$~g, $\Delta Mg\simeq 1.2$~N, and at an immersion depth $h=46$~mm.}
    \label{Fig04}
\end{figure*}

We now focus on the motion of the intruder in a granular medium whose surface is loaded with a mass $m_o=223$~g uniformly distributed over the area $w_o\ell_o$ at the center of the container. As a representative case, we present in Fig.~\ref{Fig04} the data for an intruder at depth $h=46$~mm, driven by a force $\Delta Mg\simeq 1.2$~N. Except for the applied surface overload, the experimental conditions are identical to those of Sec.~\ref{subsection_without_overload}, and results for other depths show the same qualitative behaviors. 

\begin{figure}[t]
    \centering
    \includegraphics[width=\columnwidth]{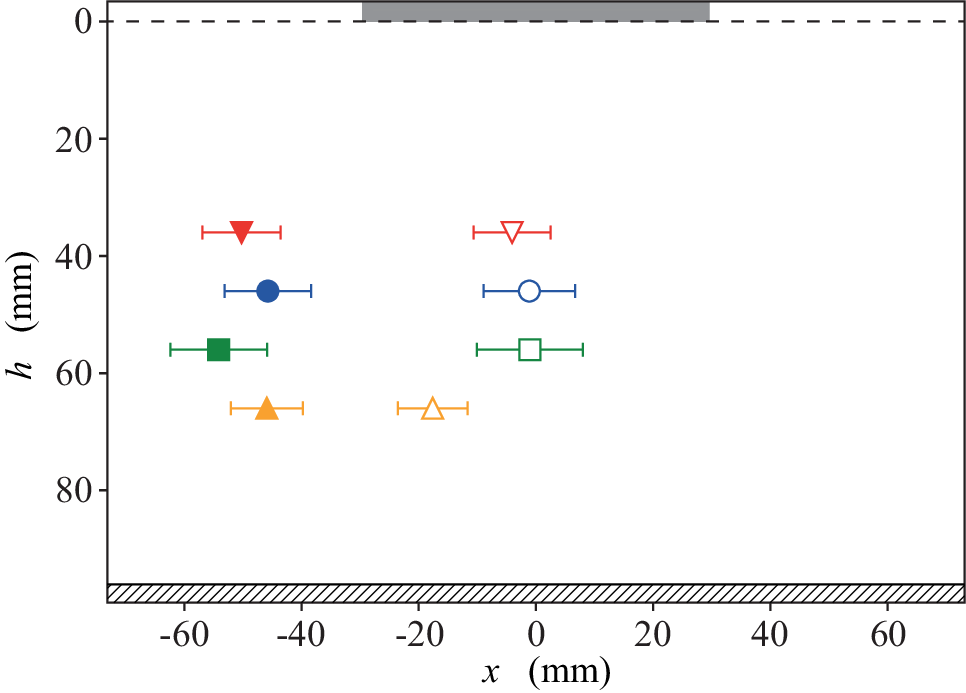}
    \caption{Region of influence of the applied surface overload of mass $m_o=223$~g for four intruder immersion depths $h$ from 36 to 66~mm. Filled symbols indicate the position where deceleration begins, while open symbols mark the onset of reacceleration. The overload is depicted by the gray area located at the granular free surface, shown as the dashed line at $h=0$. The hatched region corresponds to the bottom wall of the container at $h=96$~mm.}
    \label{Fig05}
\end{figure}

Figures~\ref{Fig04}(a) and \ref{Fig04}(b) show the position $z$ and velocity $v$ of the mass $M_1$ as a function of time, respectively. Beyond the initial transient phase at the release of mass $M_1$ (as discussed above), the surface load clearly influences the intruder dynamics. Figure~\ref{Fig04}(a) shows that the intruder position does not increase continuously with time but instead reaches a plateau while passing beneath the surface load, before recovering its accelerating motion afterward. Figure~\ref{Fig04}(b) shows the intruder velocity as a function of time for a single example, for the sake of clarity. After an initial linear rise to $v_\mathrm{max}$ (filled symbol), the velocity drops sharply to $v_\mathrm{min}$ (open symbol) before increasing again. The velocity–position plot in Fig.~\ref{Fig04}(c), based on five repetitions conducted under identical conditions, reveals a clear pattern: the intruder’s velocity initially increases, reaching a peak, at a velocity $v_\mathrm{max}$, after approximately 125~mm, followed by a sharp decline to a minimum ($v_\mathrm{min}$) around the midpoint of the trajectory, and then increases again until the intruder comes to rest against the far wall of the container. We first focus on the spatial extent of the zone where the surface load affects the intruder dynamics, defined as the region between the local maximum and minimum of the velocity. The influence zone of the overload is shown in Fig.~\ref{Fig05} for four different immersion depths $h$ of the intruder. It is observed that the influence zone of the overload is not located directly beneath it but is shifted upstream. The intruder begins to experience the effect of the overload before reaching its vertical projection. This forward shift can be explained by the fact that the intruder moves toward a region of higher confining stress where the granular material has redirected the surface load. Consequently, the intruder moves into a region with higher stress and where redirected granular forces are opposed to its motion. It is worth noting that, within the explored range of depths, the boundaries of the influence zone appear to be approximately vertical. In the case of a point load applied to a 2D assembly of photoelastic disks, it has been observed that the surface load affects a zone that widens with depth within the material \cite{reydellet2001green,geng2003green}. These findings indicate that the characteristics of the influence zone generated by a point load in a 2D disk assembly differ from those produced by an extended surface load in a 3D configuration.

\begin{figure*}[t]
    \centering
    \includegraphics[width=\textwidth]{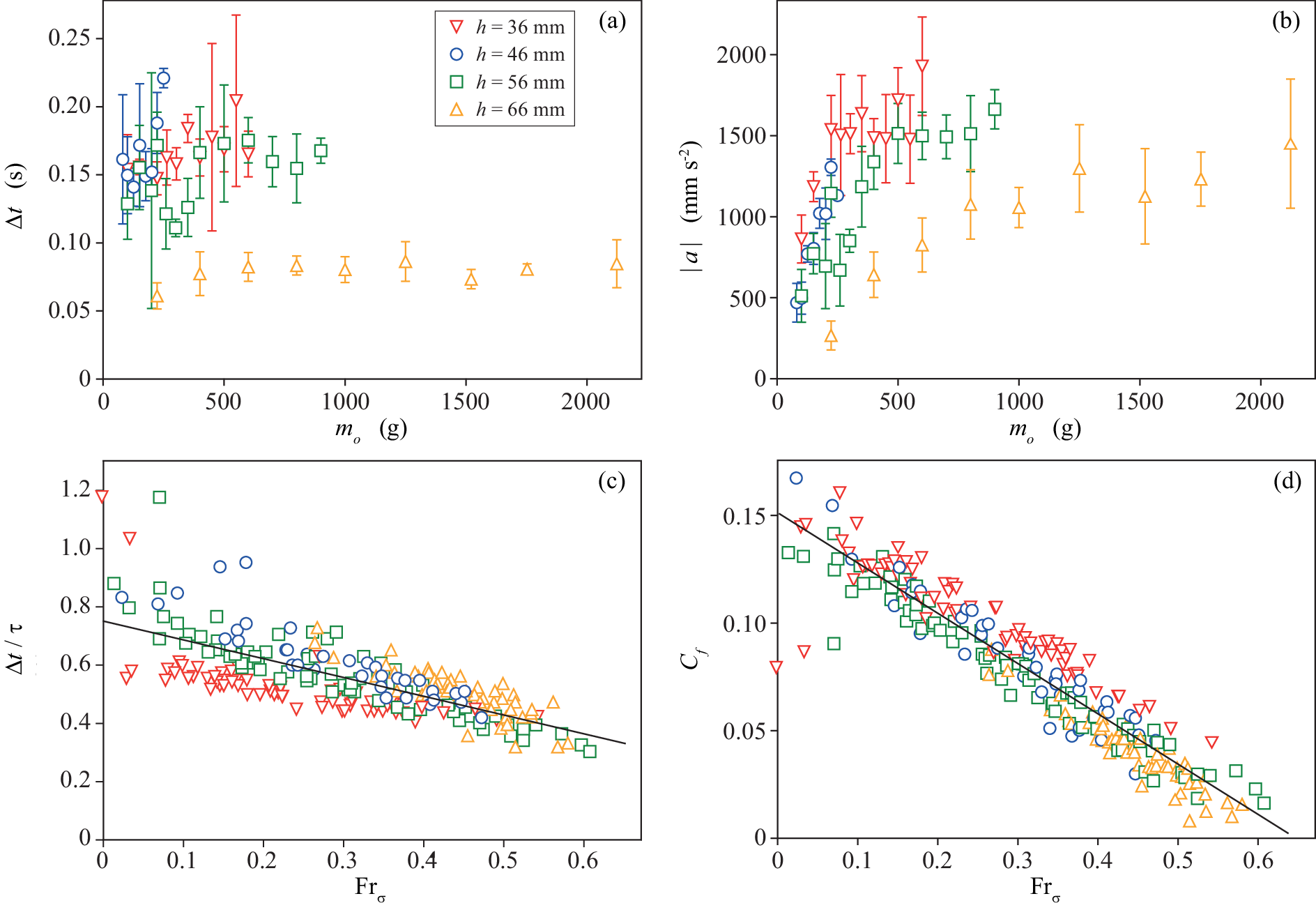}
    \caption{(a)~Duration of the deceleration $\Delta t$ and (b)~deceleration $|a|$ of the intruder as a function of the mass $m_o$ of the overload. (c)~Duration $\Delta t$ normalized by the characteristic time $\tau$ as a function of the Froude number $\mathrm{Fr}_\sigma$. The solid line corresponds to a linear fit of the data, of the form $\Delta t/\tau=-0.64\,\mathrm{Fr}_\sigma+0.75$. (d)~Overfriction coefficient $C_f$ (normalized deceleration) as a function of the Froude number $\mathrm{Fr}_\sigma$. The solid line corresponds to a linear fit of the data, of the form $C_f=-0.23\,\mathrm{Fr}_\sigma+0.15$. The different symbols correspond to different immersion depths $h$ ranging from 36 to 66~mm.}
    \label{Fig06}
\end{figure*}

Next, we examine the duration and amplitude of the intruder’s deceleration caused by the surface load. The duration is defined as the time interval $\Delta t$ during which the intruder’s velocity decreases from its local maximum to its local minimum value, $v_\mathrm{max}$ and $v_\mathrm{min}$, respectively, as displayed in Fig.~\ref{Fig04}(b). The mean deceleration experienced by the intruder due to the presence of a surface load is defined as $a= (v_\mathrm{min} - v_\mathrm{max})/ \Delta t$. For each experiment performed with varying surface loads and depths, we calculate both $\Delta t$ and $a$. Figures~\ref{Fig06}(a) and \ref{Fig06}(b) present these quantities as functions of the overload mass $m_o$. In the plots, the symbols denote mean values, while the vertical bars indicate standard deviations estimated from repeated experiments. For the time interval $\Delta t$, we observe a slight increase with $m_o$ at a given depth, while no clear trend regarding the effect of depth on this quantity can be inferred. Concerning the local deceleration induced by the surface load, we observe in Fig.~\ref{Fig06}(b) that $|a|$ first increases with $m_o$ before reaching saturation. Moreover, the deeper the intruder, the smaller the deceleration it experiences, and the saturation occurs at larger overload masses.

The influence of the intruder's depth on $\Delta t$ and $a$, as shown in Figs.~\ref{Fig06}(a) and \ref{Fig06}(b), is difficult to interpret because the driving force differs at each depth. Moreover, the local deceleration $a$ depends on the intruder’s velocity $v_{\mathrm{max}}$ immediately before the onset of deceleration, making direct comparisons between runs difficult. To address this issue, we seek dimensionless quantities relevant to the problem. The timescale $\Delta t$ can naturally be normalized by the characteristic time $\tau$ introduced earlier, yielding the nondimensional parameter $\Delta t/\tau$. For the local deceleration $a$ and, by analogy with the drag coefficient in fluid mechanics, we introduce the coefficient of overfriction, defined as
\begin{equation}
    C_f = \frac{|a|\, d}{v_\mathrm{max}^2},
    \label{eq:cf}
\end{equation}
where $v_\mathrm{max}$ is the intruder's velocity just before the deceleration occurs, as defined above. As in fluid mechanics, $C_f$ represents a dimensionless force. Finally, the intruder velocity $v_\mathrm{max}$ upon entering the zone of influence of the surface load can be normalized by introducing a Froude number, defined as
\begin{equation}
\mathrm{Fr}_\sigma = \frac{v_\mathrm{max}}{\sqrt{(\sigma_o+\rho gh)/\rho}},
\label{eq:Frp}
\end{equation}
where $\sigma_o+\rho gh$ is the granular vertical stress taking the overload $\sigma_o$ into account and $\rho$ the bulk density of the granular material. All experimental data are presented in dimensionless form, with $\Delta t$ and $C_f$ plotted as functions of $\mathrm{Fr}_\sigma$ in Figs.~\ref{Fig06}(c) and \ref{Fig06}(d), respectively. Here, the normalized time interval $\Delta t / \tau$ has been estimated using $K_z \simeq 10$, in agreement with the estimates found above (see inset of Fig.~\ref{Fig03}). Figures~\ref{Fig06}(c) and \ref{Fig06}(d) show that, when expressed in terms of nondimensional parameters, all data collapse onto a single trend.
We note that values of $\mathrm{Fr}_\sigma$ are smaller than 1, corroborating the quasistatic motion of grains, and that values for $\mathrm{Fr}_\sigma < 0.1$ deviate from the masterline because of the higher uncertainties in computing $a$ when velocity variations are very small. The collapse of the $\Delta t/\tau$ data onto a master curve corroborates $\tau$ as the relevant timescale for the problem. Moreover, the decrease of $\Delta t/\tau$ with increasing $v_\mathrm{max}$ and decreasing $\sigma_o+\rho gh$ is consistent with expectations. In the quasistatic regime, where friction is velocity-independent, the traversal time over a given distance decreases with increasing intruder velocity, consistent with the observed decrease of $\Delta t/\tau$ with $\mathrm{Fr}_\sigma$. In the same way, the collapse of the $C_f$ data when plotted in Fig. \ref{Fig06}(d) as a function of $\mathrm{Fr}_\sigma$ shows the pertinence of $v_\mathrm{max}$ and $\sigma_o+\rho gh$ for the level of deceleration. We further observe in Fig.~\ref{Fig06}(d) that $C_f$ remains within the range [0; 0.15], which can be rationalized from the geometric parameters of the experiment. A characteristic scale for the deceleration $|a|$ can be estimated by considering a velocity variation over the characteristic length of the overloaded region $w_o$, leading to $|a| = (v_\mathrm{max}^2 - v_\mathrm{min}^2) / (2w_o)$. For $v_\mathrm{min} \ll v_\mathrm{max}$, this gives $|a| = v_\mathrm{max}^2 / (2w_o)$ and Eq.~(\ref{eq:cf}) then yields the order of magnitude estimate $C_f\simeq d/(2 w_o)$. This prediction depends only on the geometry of the system and yields $C_f \simeq 0.11$, in good agreement with the experimentally observed values.

\subsubsection{Discussion}
\label{subsubsection_model}

\begin{figure*}[t]
    \centering
    \includegraphics[width=\textwidth]{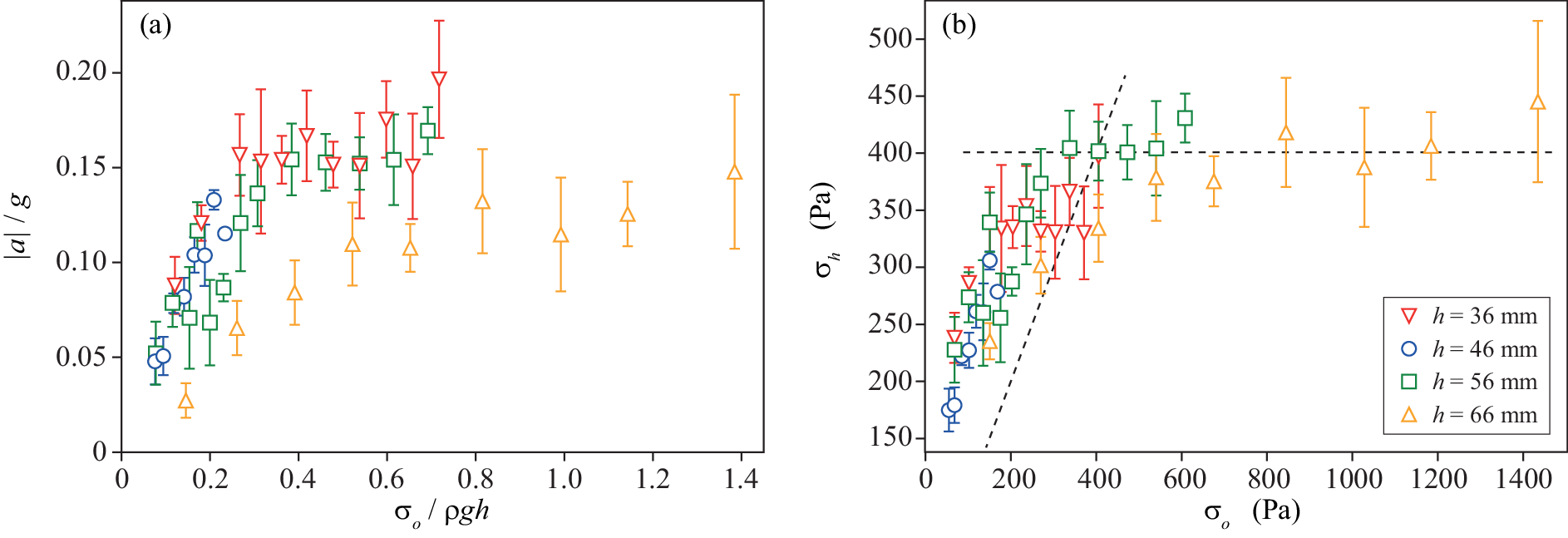}
    \caption{(a)~Normalized deceleration of the intruder $|a|/g$ as a function of the normalized surface overstress $\sigma_o/\rho gh$. (b)~Overstress $\sigma_h$ at depth $h$ as a function of the applied surface stress $\sigma_o$. The dashed lines, $\sigma_h=\sigma_o$ and $\sigma_h=$~constant, are guides to the eye. In each panel, the four symbols correspond to the different intruder depths $h$ investigated.}
    \label{Fig07}
\end{figure*}

In this section, we develop a theoretical framework to analyze the impact of a surface overload applied to the granular medium on the resulting motion of the intruder. The objective is to further analyze the deceleration experienced by the intruder as it passes beneath the overload, as shown in Fig.~\ref{Fig06}(b), in relation to $m_o$. The measured data can be represented by plotting the normalized intruder deceleration, $|a|/g$, against the normalized stress $\sigma_o /\rho gh$, for various values of the depth $h$ [Fig.~\ref{Fig07}(a)]. Similar to Fig.~\ref{Fig06}(b), we observe that $|a|$ initially increases linearly before reaching a plateau. In this case, however, all the data collapse onto a single trend, except for the largest depth, $h=66$~mm.

A stress $\sigma_o$ imposed at the surface ($h=0$) generates an additional vertical stress $\sigma_h$ at depth $h$ within the granular bed. Starting from the expression of the intruder acceleration previously established in the absence of surface loading [Eq.~(\ref{eq_a_char})], we incorporate the contribution of this overstress to determine the acceleration $a$ of the intruder as it passes beneath the overloaded region:
\begin{equation}
     a = \frac{\Delta Mg - K_z d^2 \left( \rho g h + \sigma_h \right)}{m} = a_\mathrm{qs} - \frac{ K_z d^2}{m} \sigma_h,
    \label{eq_a_dec}
\end{equation}
where, for a given depth $h$, $a_\mathrm{qs}$ denotes the intruder acceleration upstream, far from the overloaded region [Eq.~(\ref{eq_a_char})].
At this stage, it is important to clarify the nature of the stress quantity entering the drag expression. The term $\rho gh$ appearing in Eq.~(\ref{eq_a_dec}) should be understood as the vertical stress generated by the weight of the grains above the intruder, which constitutes the relevant quantity in the empirical quasistatic drag description for deeply buried objects. In this framework, the drag force is assumed to depend not on the full stress tensor, but only on this vertical stress. Accordingly, we assume that a localized surface overload modifies the drag only through an additional contribution to this same vertical stress component. This motivates the introduction of an additional vertical stress $\sigma_h$ at depth $h$, which modifies the background stress field experienced by the intruder. The local overstress $\sigma_h$ acting on the intruder at depth $h$ follows directly from the change in acceleration $(a_\mathrm{qs} - a)$, such that
\begin{equation}
     \sigma_h = \frac{m}{K_z d^2}(a_\mathrm{qs} - a).
    \label{eq_Ph}
\end{equation}
For each experiment involving an overload, we use the estimated values of $a_\mathrm{qs}$ and $a$ to determine $\sigma_h$ using Eq.~(\ref{eq_Ph}). The resulting values of $\sigma_h$ are shown in Fig.~\ref{Fig07}(b) as a function of the surface stress $\sigma_o$ applied at $h=0$. The overstress $\sigma_h$ increases with increasing $\sigma_o$ and tends to saturate at large values of $\sigma_o$. The measured values of $\sigma_h$ are comparable in magnitude to those of $\sigma_o$. For small surface stress ($\sigma_o\lesssim 400$~Pa), we find $\sigma_h\simeq \sigma_o$, which approximately reflects mechanical equilibrium in the vertical direction. In contrast, for large values of $\sigma_o$, the saturation of $\sigma_h$ suggests a redistribution of the applied stress, so that only a fraction of the surface load is transmitted to the regions explored by the intruder. Note that the transition between the linear regime and saturation appears to be independent of depth.

To interpret this behavior, we draw on the classical problem of stress transmission beneath a localized surface load. While this problem has long been studied in elasticity and soil mechanics \cite{Boussinesq, Timoshenko}, Reydellet and Clément \cite{reydellet2001green} demonstrated experimentally that when a point force $F$ is applied at the surface of a granular packing, an additional stress $\sigma$ is generated within the medium, on top of the background stress due to the weight of the grains. This extra stress is described by the Green's function \cite{Johnson1985}
\begin{equation}
    \sigma(x) = \frac{3F}{2\pi}\frac{h^3}{\left( h^2 + x^2 \right)^{5/2}},
    \label{eq:sigma_clement}
\end{equation}
where $h$ is the vertical depth and $x$ is the horizontal coordinate from the point of application of the load. Based on Eq.~(\ref{eq:sigma_clement}), we compute the infinitesimal contribution to the stress induced by a homogeneous surface stress $\sigma_o$ applied over a rectangular area of width $w_o$ and length $\ell_o$. For the sake of simplicity, we neglect variations along the $y$ direction. Denoting by $x_o$ the horizontal coordinate ranging from $-w_o/2$ and $w_o/2$, the infinitesimal stress contribution reads
\begin{equation}
    d\sigma (x(t)) = \frac{3\sigma_o\ell_o}{2\pi}\frac{h^3}{\left( h^2 + \left( x(t) - x_o \right)^2 \right)^{5/2}}dx_o,
    \label{eq:dsigma}
\end{equation}
where $(x(t)-x_o)$ is the algebraic longitudinal distance between the intruder position and the surface loading point. Incorporating this additional vertical stress into the equation of motion (\ref{eq_a_dec}), the dynamics of the intruder in the overloaded bed is governed by
\begin{equation}
    m \frac{d^2x}{dt^2} = \Delta M g - K_z d^2 \left[ \rho g h + \sigma (x(t)) \right ].
    \label{eq_motion_overloaded}
\end{equation}
Introducing dimensionless variables for length and time, this equation of motion can be rewritten as
\begin{equation}
    \frac{d^2\tilde{x}}{d\tilde{t}^2} = 1 -\Psi \int_{-\tilde{w}_0/2}^{\tilde{w}_0/2} \frac{1}{\left( 1 + \left( \tilde{x}(\tilde{t}) - \tilde{x}_0 \right)^2 \right)^{5/2}} \, d\tilde{x}_0,
    \label{eq_motion_overloaded3}
\end{equation}
where $\tilde{x}=x/h$, $\tilde{t}=t/\tau$, and $\tilde{w}_o=w_o/h$. 
Equation~(\ref{eq_motion_overloaded3}) highlights that the intruder dynamics results from the balance between the quasistatic driving acceleration (first term) and a nonlocal contribution arising from the surface overload, whose strength and spatial extent are respectively controlled by the dimensionless parameters $\Psi$ and $\tilde{w}_o$ (second term). The dimensionless parameter $\Psi$, which measures the relative importance of the surface overload, follows from the nondimensionalization using the quasistatic acceleration
$a_\mathrm{qs}$ as the natural acceleration scale:
\begin{equation}
    \Psi=\frac{3 K_z}{2\pi} \frac{\ell_o}{h} \frac{d^2 \sigma_o}{m a_\mathrm{qs}}.
    \label{eq_lambda}
\end{equation}
It is important to note that the integration bounds $\tilde{w}_o$ depend explicitly on the depth $h$. Consequently, the solution of Eq.~(\ref{eq_motion_overloaded3}) is controlled by the two parameters $\Psi$ and $\tilde{w}_o$, together with the initial conditions. The latter are given by $d\tilde{x}/d\tilde{t}=0$ and $\tilde{x}=-(L-d)/2h$ at $\tilde{t}=0$.
\begin{figure*}[t]
    \centering
    \includegraphics[width=\textwidth]{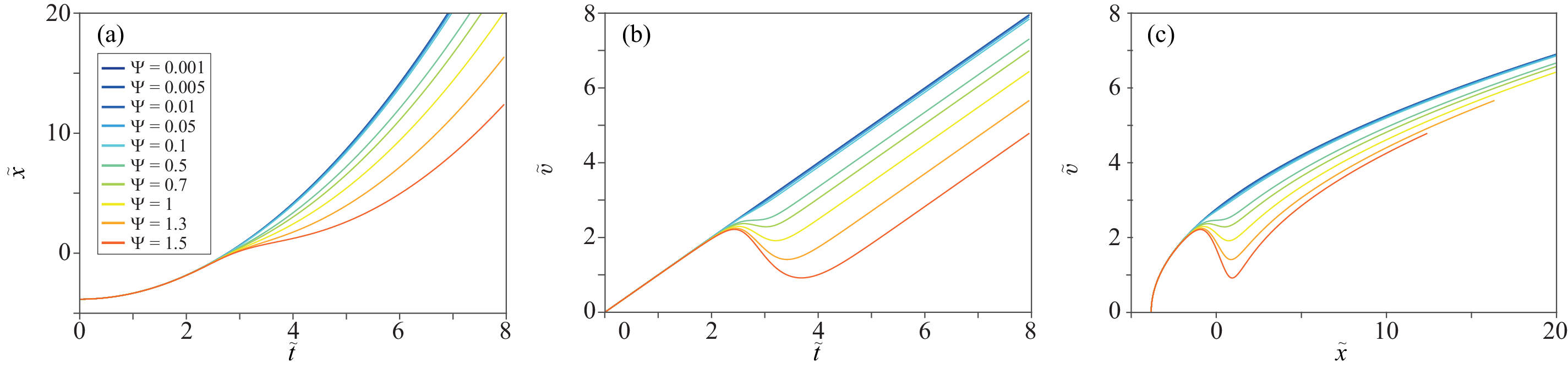}
    \caption{Time evolution of the dimensionless (a) position $\tilde{x}$ and (b) velocity $\tilde{v}$ of the intruder, and (c)~$\tilde{v}$ as a function $\tilde{x}$ for different dimensionless overload parameters $\Psi$ from 0.001 to 1.5.}
    \label{Fig08}
\end{figure*}
Equation~(\ref{eq_motion_overloaded3}) is solved numerically for different values of the dimensionless overload parameter $\Psi$. Figures~\ref{Fig08}(a) and \ref{Fig08}(b) show, respectively, the dimensionless position $\tilde{x}$ and velocity $\tilde{v}$ of the intruder for various applied overload parameters $\Psi$. The dimensionless variables are computed using parameters consistent with the experimental conditions: $h=46$~mm, $\Delta M=120.9$~g, $d=12$~mm, $m=224.4$~g, and $K_z=10$ extracted from experiments presented in the inset of Fig.~\ref{Fig03}.
The applied surface stress $\sigma_o$ ranges from 0 to 350~Pa, which is the maximal stress which can be solved numerically. Except for the initial and final stages of the motion discussed in Sec.~\ref{subsection_without_overload}, the model captures the intruder dynamics well, reproducing the deceleration observed when the intruder passes beneath the overloaded region, as previously shown in Fig.~\ref{Fig04}. Moreover, the magnitude of the deceleration increases with the applied stress. The model also predicts the existence of a critical stress above which the intruder velocity approaches zero, and the intruder eventually stops. This behavior corresponds to a limiting overload for a given driving force, in agreement with the experimental observations.

From this analysis, the normalized deceleration $\tilde{a}=a/a_\mathrm{qs}$ can be extracted from Fig.~\ref{Fig08}(b) using the same procedure as for the experimental data [Fig.~\ref{Fig04}(b)]. Using Eq.~(\ref{eq_Ph}), we estimate the effective overstress felt by the intruder in this model,
\begin{equation}
    \tilde{\sigma_h} = \frac{K_z d^2 \sigma_h}{m a_\mathrm{qs}} =  1 - \tilde{a}.
    \label{eq_Ph_tilde}
\end{equation}
Since the normalized overstress $\tilde{\sigma_h}$ can be determined both experimentally and theoretically, a quantitative comparison between the model and the experiments can be performed.
\begin{figure}[t]
    \centering
    \includegraphics[width=\columnwidth]{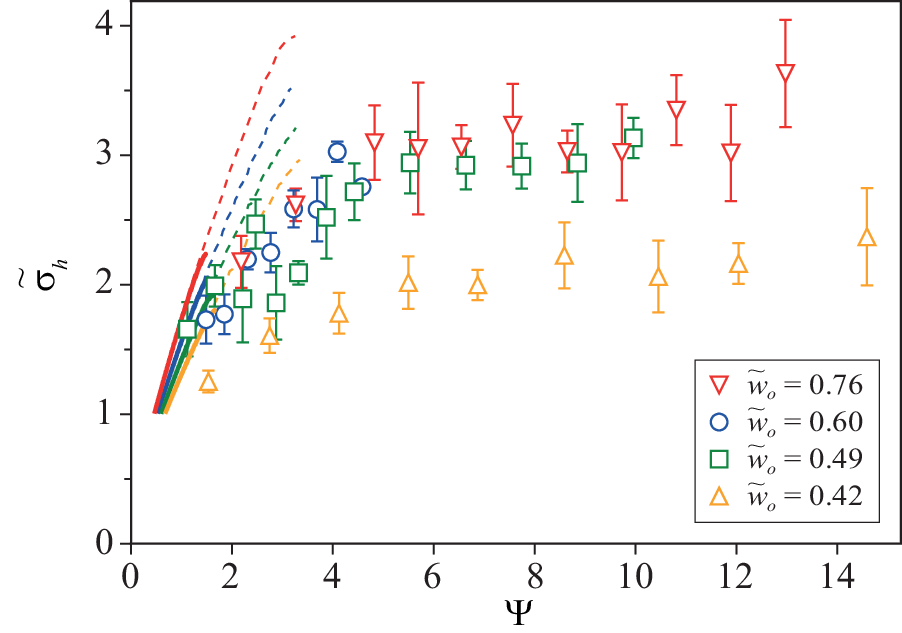}
    \caption{Normalized overstress $\tilde{\sigma_h}$ at depth $h$ as a function of $\Psi$ for different normalized width $\tilde{w}_o$ and initial condition $L=364$~mm. Solid lines correspond to the theoretical computation of $\tilde{\sigma_h}$ for the same values of $\Psi$, $\tilde{w}_o$, and $L$. Dashed lines correspond to the theoretical computation of $\tilde{\sigma_h}$ in similar conditions for $L=900$~mm.}
    \label{Fig09}
\end{figure}
Figure~\ref{Fig09} displays the evolution of $\tilde{\sigma_h}$ as a function of the dimensionless overload parameter $\Psi$ for the four investigated immersion depths. The symbols correspond to the experimental measurements, while the solid lines represent the theoretical predictions obtained by solving Eq. (10) with initial conditions matching the experimental configuration, namely, an intruder released from rest at the entrance of the container.

The model satisfactorily captures the initial linear increase of $\tilde{\sigma_h}$ with $\Psi$ without introducing any adjustable parameter beyond those independently determined from the unloaded dynamics. However, for the experimental initial conditions, the theoretical curves terminate at relatively small values of $\Psi$ ($\Psi \lesssim 2$). Beyond this critical value, the model predicts that the intruder velocity vanishes beneath the overloaded region and no longer crosses it. 
In this situation, the deceleration cannot be defined consistently with the experimental procedure, which relies on identifying a local maximum and minimum of the velocity during a complete traversal of the overloaded zone.  As a consequence, the theoretical curves obtained under experimental conditions extend, at best, over the very first points of experimental data, which limits the direct quantitative comparison over a broader range of $\Psi$.

To explore the intrinsic behavior of the model beyond this restriction, we also computed theoretical solutions using modified initial conditions. More specifically, the intruder was released from a position located at a larger distance upstream of the overloaded region ($L=900$\ mm), so that it reaches the overloaded region with a higher velocity. This increased approach velocity enables extending the accessible range of the theoretical predictions. The corresponding results are shown as dashed lines in Fig.~\ref{Fig09}.
Although these additional curves no longer correspond strictly to the experimental configuration, they provide useful insight into the structure of the theoretical response. In particular, they highlight that $\tilde{\sigma_h}$ initially increases roughly linearly with $\Psi$, before slightly bending and eventually approaching a regime associated with strong deceleration and imminent arrest.

Small discrepancies between theory and experiments remain visible even in the range where both can be compared quantitatively, with the model slightly overestimating the measured values. These differences are not unexpected, as the stress transmission is computed using a Green’s function originally derived for the response of a granular medium to a localized point force in a two-dimensional framework, and is here employed as an approximation for a three-dimensional configuration involving a spatially extended load. Effects such as stress redistribution within the bulk, frictional interactions with the lateral walls, and the finite depth of the granular layer are therefore not explicitly accounted for. In particular, Eq.~(\ref{eq:sigma_clement}) is derived for a semi-infinite medium, whereas the present experiments are performed in a granular bed of finite depth. These effects may contribute to the discrepancies observed between theory and experiments, especially for the largest immersion depth. In addition, the implementation of the model relies on two parameters, $\Psi$ and $\tilde{w}_o$, which both vary with the intruder depth $h$. Furthermore, the present model is not intended to predict the absolute value of the quasistatic drag coefficient $K_z$, which is independently determined from the unloaded experiments and subsequently kept fixed throughout the analysis of the overloaded configurations. A prediction of this reference drag coefficient would require a more elaborate description of the quasistatic stress field, such as slip-line approaches developed in the literature \cite{Kang}. Despite these limitations, once the reference drag coefficient has been independently determined, the model provides a consistent description of the initial response of the intruder to surface loading and successfully captures the main experimental trends over the accessible range of $\Psi$.

\section{Conclusion}

We have investigated experimentally and theoretically the motion of a spherical intruder driven horizontally within a granular bed subjected to a localized surface overload. In the absence of loading, the intruder undergoes a constant acceleration governed by a depth-dependent quasistatic drag. When a surface overload is applied, a significant transient deceleration occurs as the intruder passes beneath the loaded region. The magnitude of this deceleration increases with the applied stress, decreases with depth, and ultimately saturates at large overloads, revealing a partial redistribution and screening of stresses within the granular bulk. This saturation, which is not trivial {\it a priori}, indicates that beyond a certain level of surface loading, additional stress is no longer fully transmitted to the region explored by the intruder, but is instead redistributed laterally or redirected toward the container boundaries.
By introducing appropriate dimensionless quantities, namely, a characteristic timescale and an overload-based Froude number, we showed that the deceleration dynamics collapse onto master trends. A model based on stress transmission from the surface quantitatively captures the linear response corresponding to small overloads. Although simplified, the model provides a consistent framework to rationalize how localized surface stresses affect subsurface granular motion. The present study focuses on relatively deep intruders ($0.18\lesssim d/h \lesssim 0.33$), for which no significant deformation of the free surface is observed. For shallower intrusions (larger $d/h$), surface deformation and associated free-surface flows are expected to modify both the drag force and the overload-induced deceleration.

Beyond the present study, several perspectives naturally emerge. Numerical simulations would provide direct access to the internal stress distribution and force-chain organization, helping to clarify the physical origin of the observed saturation and to disentangle the respective roles of stress arching, stress redirection, and wall effects.
Extending the investigation to the inertial regime also constitutes a natural continuation of this work. At larger Froude numbers, where collisional contributions to the drag become significant, the interplay between overload-induced stresses and inertial effects may lead to different scaling behaviors and modified deceleration dynamics.
Another promising direction concerns the vertical force balance. Since surface loading is expected to affect not only the horizontal drag but also the lift force acting on the intruder, measuring and modeling this effect would provide further insight into three-dimensional stress transmission in loaded granular media.

Altogether, this work contributes to a quantitative understanding of intrusion processes in loaded granular media, and opens the way toward a broader understanding of mobility in granular systems.

\section{\label{sec:Ack} ACKNOWLEDGMENTS}
The authors thank J.~Amarni, A.~Aubertin, L.~Auffray, C.~Manquest, and R.~Pidoux for their technical support. This work has benefited from fruitful discussions with G.~Gauthier. The authors are grateful to FAPESP (Grants No. 2018/14981-7, No. 2024/13295-3, and No. 2024/13981-4) for the financial support provided. This work has also been supported by “Investissements d’Avenir” LabEx PALM (Grant No. ANR-10-LABX-0039-PALM).

\section{DATA AVAILABILITY}
All data used in this work are publicly available on an open repository \cite{Supplemental2}.
%The data that support the findings of this article are openly available \cite{Supplemental2}.

\bibliography{references}

\end{document}